\documentclass{article}
\usepackage{spconf,amsmath,graphicx}
\usepackage{booktabs}
\usepackage{multirow}
\usepackage{makecell}
\usepackage{rotating}
\usepackage{caption} 
\usepackage{amsfonts}
\usepackage{array}
\usepackage{url}
\usepackage{enumitem}

\title{Speech foundation models on intelligibility prediction \\ for hearing-impaired listeners}
\name{Santiago Cuervo,\quad Ricard Marxer \thanks{%
The authors thank the French National Research Agency for their support through the ANR-20-CE23-0012-01 (MIM) grant, and the Institute of Convergence ILCB, supported by grants from France 2030 (ANR-16-CONV-0002) and the Excellence Initiative of Aix-Marseille University (A*MIDEX).}}

\address{Universit\'e de Toulon, Aix Marseille Univ, CNRS, LIS, France}
\begin{document}
\maketitle
\begin{abstract}
\textit{Speech foundation models} (SFMs) have been benchmarked on many speech processing tasks, often achieving state-of-the-art performance with minimal adaptation. However, the SFM paradigm has been significantly less explored for applications of interest to the speech perception community. In this paper we present a systematic evaluation of 10 SFMs on one such application: Speech intelligibility prediction. We focus on the non-intrusive setup of the \textit{Clarity Prediction Challenge 2} (CPC2), where the task is to predict the percentage of words correctly perceived by hearing-impaired listeners from speech-in-noise recordings. We propose a simple method that learns a lightweight specialized prediction head on top of frozen SFMs to approach the problem. Our results reveal statistically significant differences in performance across SFMs. Our method resulted in the winning submission in the CPC2, demonstrating its promise for speech perception applications.

\end{abstract}
\begin{keywords}
Foundation models, speech perception, intelligibility prediction, hearing aids.

\end{keywords}
%


\section{Introduction}
\label{sec:intro}

Foundation models (FMs) are deep neural networks trained on large diverse datasets that can be applied to a wide range of downstream tasks with little or no adaptation
. 
They have led to great progress in natural language processing problems on both text and speech \cite{devlin2018bert, baevski2020wav2vec, hsu2021hubert, radford2021speech, chen2021wavlm}, often reaching state-of-the-art performance with little task-specific training. 

In the speech domain, which is the focus of our work, several studies have benchmarked speech FMs (SFMs) on their capacity to extract phonetic, semantic, speaker-related and paralinguistic information \cite{yang2021superb, nguyen2020zero}. However, there is a lack of systematic benchmarks on speech perception tasks, in which the objective is not to decode the content of the speech signal, but to model how the signal will be decoded by a listener.
 
We believe that SFMs have significant potential to drive progress on speech perception. Our belief is based on two facts: \textit{a)} Representations learned by SFMs can capture semantically meaningful information from both speech \cite{yang2021superb} and the acoustic background in which it occurs \cite{gong23whisperat}, and \textit{b)} Language processing in SFMs and brains shows significant similarities \cite{millet22speech}. Based on \textit{a)} we deduce that SFMs can extract the listener-independent information relevant for perception from the whole soundscape. From \textit{b)}, we speculate that such similarities translate in predictive power over perceptual phenomena. We presented evidence for this hypothesis in \cite{cuervo23_interspeech}.

In this work we take a step towards benchmarking SFMs for speech perception. We systematically evaluate 10 different SFMs on the \textit{Clarity Prediction Challenge 2} (CPC2) \cite{graetzer-clarity}, in which the objective is to build machine learning models that take speech-in-noise binaural audio recordings and the audiogram from a hearing-impaired listener as input, and predict the percentage of words that the listener correctly recognizes. Our main contributions are:

\begin{itemize}
    \setlength{\itemsep}{0pt}
    \item We present a prediction model for the CPC2 leveraging SFMs. The model uses a specialized prediction head that extracts multi-layer features from SFMs, and uses cross-attention to exploit binaural cues. This method resulted in the winning submission for the CPC2.
    \item We benchmark 10 SFMs on the CPC2 using our specialized predictive model, finding statistically significant performance differences. Additionally, we study SFM ensembles obtained by averaging predictions. We demonstrate that SFMs learn complementary information, as evidenced by over 60\% of their ensembles outperforming the best individual model, and all ensembles outperforming all but the single best model.
\end{itemize}

\let\thefootnote\relax\footnotetext{© 2024 IEEE. Personal use of this material is permitted. Permission from IEEE must be obtained for all other uses, in any current or future media, including reprinting/republishing this material for advertising or promotional purposes, creating new collective works, for resale or redistribution to servers or lists, or reuse of any copyrighted component of this work in other works.}

\section{Intelligibility prediction model}

\begin{figure*}[t!]
\setlength{\belowcaptionskip}{-10pt}
  \centering \includegraphics[width=0.74\textwidth]{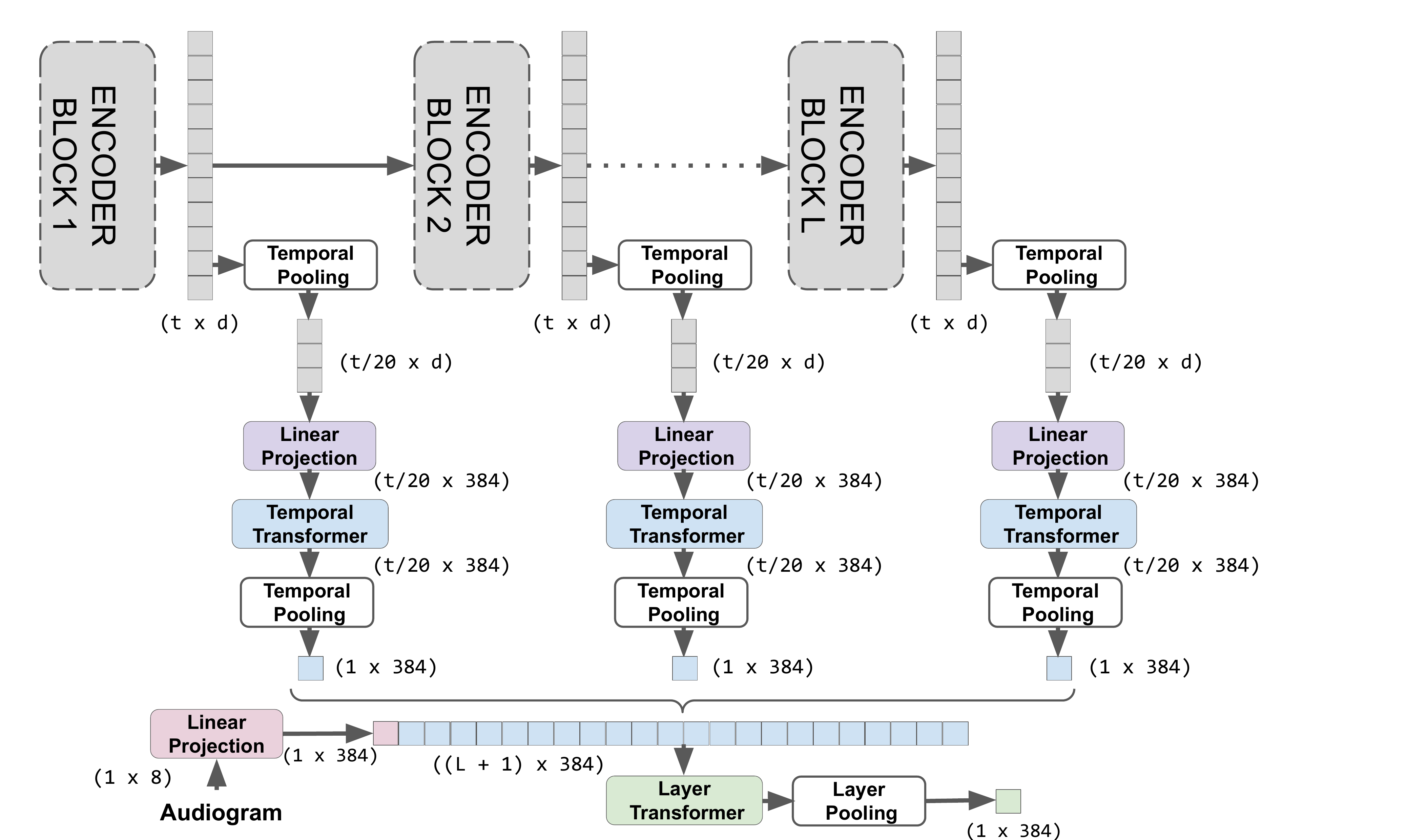}
  \centering \includegraphics[width=0.25\textwidth]{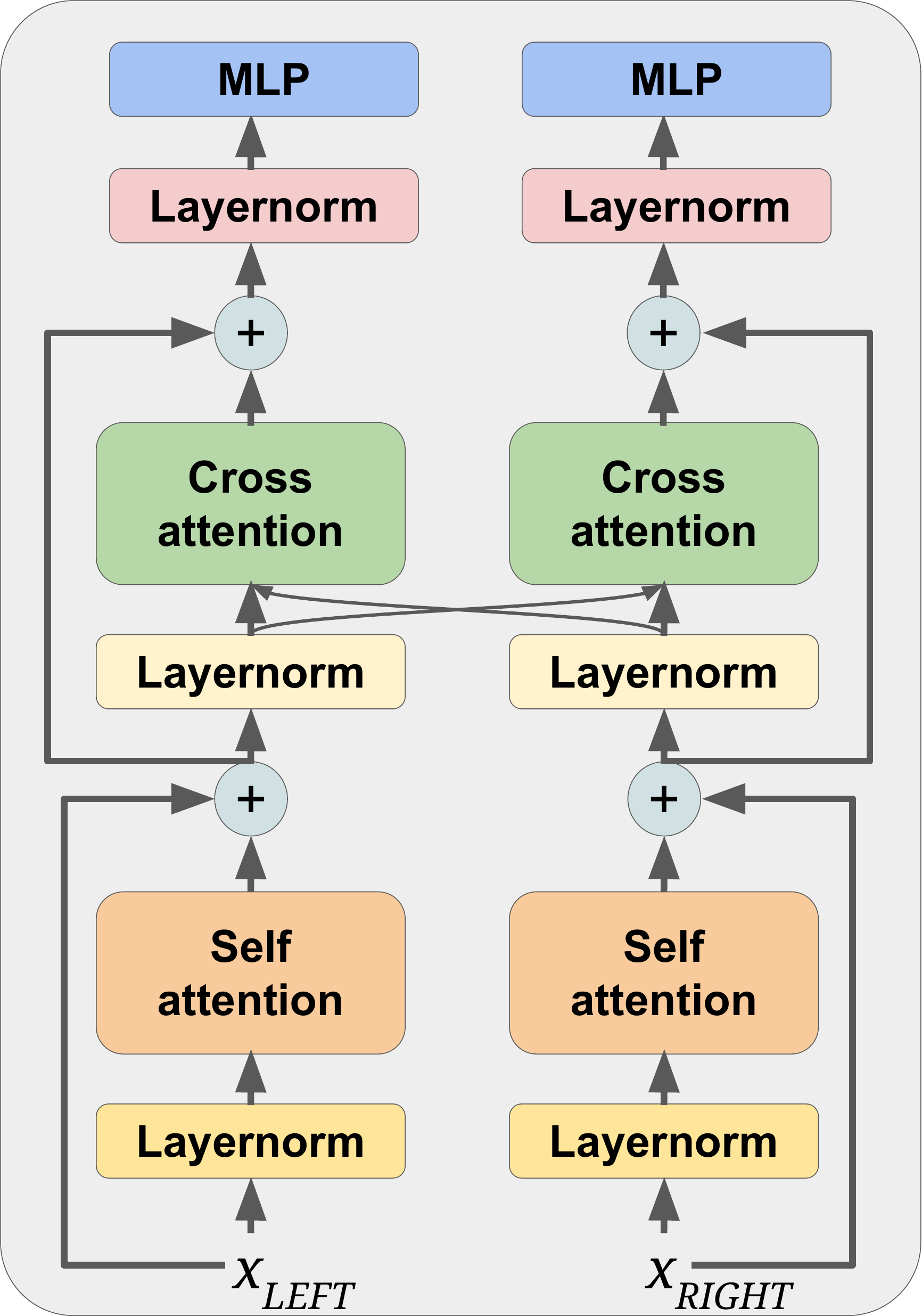}
  \caption{Intelligibility prediction model architecture. With the exception of frozen backbone blocks (grey), blocks with the same color indicate shared parameters. \textit{Left:} Pipeline applied to each channel of the binaural signal
  . \textit{Right:} Binaural block used in the temporal and layer transformers. The cross-attention layer enables modeling of non-linear binaural interactions.
  }
\label{fig:diag}
\end{figure*}

Our model for intelligibility prediction is depicted in Fig. \ref{fig:diag}. It consists of a pre-trained frozen backbone that extracts representations from raw binaural speech data, and a head designed to predict intelligibility conditioned on the representations obtained from the backbone. We extract features from all layers of the backbone following \cite{yang2021superb, chen2021wavlm, gong23whisperat}.

\subsection{Backbone}

For the backbone we consider multiple state-of-the-art SFMs:

\textbf{Wav2vec 2.0} consists of two components: An encoder and a predictor. The encoder is a strided convolutional network that produces embeddings from the speech input. The predictor is an encoder-only transformer \cite{vaswani17transformer} trained contrastively to distinguish positive and negative samples from discretized and masked segments of the encoder’s output. We use four variants: \textsc{Large} pretrained wav2vec 2.0 \cite{baevski2020wav2vec} trained on 60k hours of the LibriLight dataset (LL-60k), \textsc{Large} pretrained wav2vec 2.0 robust \cite{hsu21_interspeech} trained on 3k hours of additional diverse data, and their variants fine-tuned (FT) for ASR on 960h of the LibriSpeech dataset (LS-960h).

\textbf{HuBERT} uses an encoder-predictor architecture as in wav2vec 2.0. Instead of a contrastive loss, it is trained with a masked language modeling task similar to BERT \cite{devlin2018bert}, but masks continuous segments of the encoder's output instead of tokens. The targets for the masked prediction are obtained through unsupervised clustering of raw speech features or learned features from earlier iterations of the model. We use the \textsc{Large} and \textsc{X-Large} variants \cite{hsu2021hubert} pre-trained on LL-60K, and their versions fine-tuned for ASR on LS-960h.

\textbf{WavLM} expands on HuBERT with an auxiliary denoising task. Some inputs are corrupted by mixing them with noise or other speech. The model is trained to jointly predict the targets for the masked segments and original uncorrupted speech. This joint pre-training helps the model learn useful representations for a wide range of speech tasks. We use the \textsc{Large} variant \cite{chen2021wavlm} pre-trained on 94k hours of speech.

\textbf{Whisper} is trained through large-scale multi-task supervised learning on over 680k hours of diverse audio, unlike previous models which were trained using self-supervised learning. The model is an encoder-decoder transformer trained on many speech processing tasks, including multilingual speech recognition, speech translation, spoken language identification, and voice activity detection. Whisper showed robustness to unseen domains, being often competitive with prior fully supervised results but in a zero-shot setting. We use the encoder of the \textsc{Large}-v2 variant \cite{radford2021speech}.

\subsection{Prediction head}

We adapt the transformer across time and layers from \cite{gong23whisperat} for intelligibility prediction by adding an embedding layer to insert listener information, and a cross-attention block between binaural channels to account for binaural effects on intelligibility. We describe below its components.

\textbf{Dimensionality reduction}: The output of the backbone is an $L \times t \times d$ tensor, $L$ being the number of layers of the SFM, $t$ the length in time of the speech representations sequence, and $d$ the dimension of the representations. Considering the quadratic computational complexity of transformers on the sequence length, we downsample in time by a factor of 20 through average pooling, following \cite{gong23whisperat}. We then project the embeddings from their original $d$-dimensional space to a $384$-dimensional space through a learnable linear projection.

\textbf{Temporal attention pooling}: We use a transformer with bidirectional attention to compress the downsampled sequence along the time axis. As in \cite{devlin2018bert}, we append a special \texttt{CLS} embedding at the beginning of the sequence, which after being processed by the transformer can be used as a compressed representation of the whole sequence. Temporal pooling gives an $L \times 384$ tensor.

\textbf{Audiogram embedding}: We use audiograms as listener's information. Audiograms are a vector in $\mathbb{R}^{+8}$ representing hearing capacity at 8 different frequencies. We project it to $\mathbb{R}^{384}$ through a learnable linear projection and append it along the layer axis, yielding an $(L + 1) \times 384$-dimensional tensor.

\textbf{Attention pooling across layers}:  Next, we apply attention pooling along the layer axis using the same method as in temporal attention. We use the special \texttt{CLS} embedding to summarize the sequence of layer and audiogram embeddings, resulting in a single $384$-dimensional representation.

\textbf{Binaural cross-attention}: Each binaural channel is processed by the pipeline described above. To account for non-linear binaural effects in intelligibility \cite{grange16hrintelligibility} we use cross-attention in the attention pooling transformers (see Fig.~\ref{fig:diag}). 

\textbf{Prediction layer}: The whole process yields two $384$-dimensional representations, one per binaural channel. Both representations are averaged and linearly projected to $\mathbb{R}$. We apply a sigmoid layer to the output and multiply by 100 to obtain predictions in the range of the target.

\section{Experiments and Results}

\subsection{Experimental setup}

\subsubsection{Data, metrics, and baselines}

    CPC2 provides binaural speech signals processed by hearing aids, audiograms from hearing impaired listeners to which the signals are presented, and their responses. The data is split into three \texttt{train}/\texttt{test} partitions. The challenge requires training three prediction systems, one per train set. There are 24,630 train and 897 test samples. For each partition, we sample a \texttt{dev} set using complex scenes from another partition. Signals are at 32 kHz, we resampled to 16 kHz for the SFMs.

We report average \texttt{test} RMSE, the metric used in the CPC2. Significance is tested with a paired Wilcoxon signed-rank test \cite{wilcoxon45}. Baselines are logistic regression on HASPI \cite{kates14haspi} (CPC2 baseline) and CPC2 SFM-based submissions.

\subsubsection{Training}

All models were trained to minimize a Huber loss for 60k steps using Adam \cite{kingma15adam} with a learning rate of $3\mathrm{e}{-5}$, $\beta_1 = 0.9$, $\beta_2 = 0.98$, and a batch size of 160. We used a cosine learning rate schedule with a linear warm-up of 2000 steps. To all transformer layers we apply dropout \cite{srivastava14dropout} with $p=0.1$.

\subsubsection{Software and computational cost}

We use publicly available checkpoints for the backbones \footnote{\url{https://github.com/facebookresearch/fairseq/}, \url{https://github.com/microsoft/unilm}, and \url{https://github.com/openai/whisper}}.
A training run takes close to 9 hours on a single GPU NVIDIA A100-80 GB and it requires about 18.4 GB of GPU memory when using HuBERT \textsc{X-Large} or Whisper, and 14.4 GB with other models. 
We will release our source code at \url{https://github.com/tiagoCuervo/sfms-cpc2}.

\subsection{Results}

\subsubsection{Backbone performance}

\begin{table}[t!]
\caption{Prediction \texttt{test} RMSE on CPC2 for different backbones using our model. \textit{Top:} across 15 independent runs and of best models chosen by cross validation. \textit{Bottom:} for our submission and baselines.}
\label{tab:results}
\setlength{\tabcolsep}{5pt}
\centering
\setlength\extrarowheight{-3pt}

\begin{tabular}{lcc}
\toprule
& \multicolumn{2}{c}{\texttt{test} RMSE $\downarrow$} \\
\multicolumn{1}{c}{Backbone} &      Min / Mean / Max &  \makecell{Best \\ on \texttt{dev}} \\
\midrule
 wav2vec 2.0 & 27.15 / 28.64 / 30.15 & 28.27 \\
 wav2vec 2.0 FT & 26.65 / \textbf{27.76} / \textbf{28.80} & 26.74 \\
 wav2vec 2.0 robust & 27.21 / 29.21 / 30.31 & 28.74 \\
 wav2vec 2.0 robust FT & 26.75 / 30.05 / 32.39 & 27.36 \\
 HuBERT L & \textbf{25.05} / 27.89 / 29.52 & \textbf{25.05} \\
 HuBERT L FT & 26.24 / 27.94 / 30.09 & 26.92 \\
 HuBERT XL & 26.69 / 29.47 / 32.80 & 26.85 \\
 HuBERT XL FT & 27.40 / 30.61 / 33.50 & 28.50 \\
 WavLM & 25.28 / 27.88 / 29.03 & 25.28 \\
 Whisper & 26.23 / 28.85 / 30.73 & 27.83 \\ \midrule
  \makecell[l]{CPC2 baseline} (HASPI) & 28.70 / - / - & - \\
 \makecell[l]{CPC2 E023 (Whisper)} & 26.34 / - / - & - \\
 \makecell[l]{CPC2 E002 (Whisper)} & 25.30 / - / - & - \\
  CPC2 E011 (ours) & 25.10 / - / - & 25.10 \\
\bottomrule
\end{tabular}
\end{table}

\begin{table}[t!]
    \caption{Model ranking according to statistical significance with respect to wav2vec 2.0 FT. $P$-values for a Wilcoxon paired test on the hypothesis of model$_i$ having greater RMSE.
    }
    \label{tab:ranking}
    \centering
    \setlength\extrarowheight{-3pt}
    \setlength{\tabcolsep}{2pt}
    \begin{tabular}{lc}
    \toprule
                          model$_i$ &     \makecell{Wilcoxon $p$-val \\ RMSE$_i$ $>$ RMSE$_{\text{wav2vec 2.0 FT}}$} \\
    \midrule
                            WavLM & \textbf{0.598022} \\
          HuBERT L FT & \textbf{0.319336} \\
                     HuBERT L & \textbf{0.151398} \\
                     HuBERT XL & 0.047688 \\
                      wav2vec 2.0 & 0.037302 \\
               wav2vec 2.0 robust & 0.037302 \\
                          Whisper & 0.036499 \\
    wav2vec 2.0 robust FT & 0.027679 \\
      HuBERT XL FT & 0.006226 \\
    \bottomrule
    \end{tabular}
\end{table}

Table \ref{tab:results} shows results for different SFM backbones.
All backbones outperform the CPC2 baseline. Wav2vec 2.0 FT has the lowest average RMSE, but HuBERT L has the overall lowest. Three other models outperform the best wav2vec 2.0 FT. Statistical tests (Table \ref{tab:ranking}) show wav2vec 2.0 FT does not significantly outperform WavLM, Hubert L FT, and Hubert L. These four can be considered the best performing models.

\subsubsection{Binaural cross-attention ablation}

We evaluated the influence of binaural cross-attention in our predictive model. We conducted ablations with two backbones with significant statistical differences: WavLM and Whisper. The results are presented in Table \ref{tab:binaural}, and show that binaural cross-attention significantly improves performance.

\subsubsection{Our submission to the CPC2}

Our submission was an ensemble between the models with the lowest \texttt{dev} RMSE using WavLM and Whisper as backbones. The results we obtained are shown in Table \ref{tab:results}. Although our submission was an ensemble, the results show that predictions of single WavLM or HuBERT L backbones would have outperformed others.  

\begin{table}[t!]
\caption{
Effect of binaural cross-attention on performance measured across 15 independent runs.
}
\label{tab:binaural}
\centering
\setlength{\tabcolsep}{4pt}
\setlength\extrarowheight{-3pt}
\begin{tabular}{lccc}
\toprule
& \multicolumn{2}{c}{\texttt{test} RMSE $\downarrow$} & \makecell{Wilcoxon $p$-val} \\
                       Model & \makecell{with Binaural \\ attention}                                                & \makecell{w./out Binaural \\ attention}                                            & \makecell{RMSE$_{\text{with}}$ \\ $<$ RMSE$_{\text{w/out}}$}                                                                                                                                            \\ \midrule
WavLM                  & \textbf{27.88} $\pm$ 2.28 & 28.65  $\pm$  1.94  & 0.021322                                                                                                                                    \\
Whisper                & \textbf{28.85} $\pm$ 2.64                                                     & 29.46 $\pm$ 2.07                                                     & 0.018101 \\ \bottomrule
\end{tabular}
\end{table}

\subsubsection{Ensemble performance}

From our submission we learned that using ensembles of predictions from different backbones improved performance. Therefore, we experimented using ensembles obtained by averaging the predictions from the best performing models for each backbone, chosen by cross validation. Results for the two best and worst ensembles are presented in Table \ref{tab:ensembles}, showing that a relative improvement of 5\% could be obtained using the best ensemble. It also should be noted that even the worst ensemble outperforms all but the best single model. Furthermore, the whole set of results showed that over 60\% of ensembles outperform the best single model.

\begin{table}[]
\caption{Two best (top) and worst (bottom) performing ensembles.
}
\label{tab:ensembles}
\centering
\setlength{\tabcolsep}{2pt}
\setlength\extrarowheight{-3pt}
\begin{tabular}{llc}
\toprule
Rank &                        \makecell[c]{Ensemble} &      RMSE $\downarrow$\\
\midrule
1 & HuBERT L+wav2vec 2.0 robust FT & 23.86 \\
2 &                HuBERT L+WavLM & 23.88 \\
\midrule
44 &            wav2vec 2.0 robust+Hubert XL & 25.24 \\
45 & wav2vec 2.0 robust+wav2vec 2.0 robust FT & 25.25 \\
\bottomrule
\end{tabular}
\end{table}

\section{Related work}

Our predictive model was first presented in our submission to the CPC2. With respect to that work, we significantly expanded the amount of experiments, including statistically validated benchmarking of 8 additional SFMs, ablations of binaural cross-attention, and evaluation of SFMs ensembles. Moreover, the results here presented outperform our submission and refute the main hypothesis on which we based it (see section \ref{sec:discussion}). Our paper is also related to previous works benchmarking SFMs on the CPC1 \cite{close2023non, edozezario22_interspeech}, but goes beyond them by considering a larger set of SFMs and showing state-of-the-art performance in the most recent edition of the challenge. 

\section{Discussion}
\label{sec:discussion}
Our results showed that four SFMs performed significantly better than others. All but one of them (wav2vec 2.0 FT) use masked language modeling training, so this could be a factor contributing to their performance. Three out of four of the best models were trained on clean speech only. This refutes the hypothesis behind our CPC2 submission, namely that backbones trained on speech with diverse backgrounds should perform best, and which led us to choose WavLM and Whisper for our submission. Contrary to the trend in other applications, the largest models (Hubert XL and Whisper) performed worse. This could be due to overfitting to extra variance captured by models with higher capacity. 

Our experiments with ensembles demonstrated that different SFMs can compensate for each other biases. This suggests that a SFM better suited for intelligibility prediction could be achieved by combining properties from the models studied.

Regarding the design choices for our prediction head, more experiments could be needed to justify them. Binaural cross-attention was shown to have significant impact in performance, however further experiments are required to determine if this is due to exploiting binaural cues, or to gains in expressive power from the extra parameters in the cross-attention block. The effects of downsampling should also be studied, as it could obfuscate temporal fine-structure useful for intelligibility prediction. The role of the audiogram in the models' predictions remains an interesting question to be studied in future work.

\section{Conclusions}

We presented a prediction model leveraging SFMs that obtains state-of-the-art performance in the CPC2. Using that predictive model we benchmarked 10 different SFMS, revealing statistically significant differences in performance. Our experiments also showed that ensemble methods show meaningful improvements, and suggest that different SFMs learn different information useful for intelligibility prediction on hearing-impaired listeners.

\bibliographystyle{IEEEbib-abbrev}
\bibliography{refs}

\end{document}